\documentstyle[aps,prb,subfigure,epsfig,amsmath,amssymb]{revtex}
\begin{document}
\draft

\title{Analytical expressions for the charge-charge 
local-field factor and
the exchange-correlation kernel of
a two-dimensional electron gas}
\author{B. Davoudi$^{1,2}$, M. Polini$^1$, G. F. Giuliani$^3$ 
and M. P. Tosi$^1$}
\address{$^1$NEST-INFM and 
Classe di Scienze, Scuola Normale Superiore, I-56126 Pisa, Italy\\
$^2$Institute for Studies in Theoretical Physics and Mathematics, Tehran
19395-5531, Iran\\
$^3$ Physics Department, Purdue University, West Lafayette, Indiana
}
\maketitle
\vspace{0.2 cm}

\begin{abstract}
We present an analytical expression for the static many-body 
local field factor $G_{+}(q)$ of a homogeneous two-dimensional electron 
gas, which reproduces Diffusion Monte Carlo data and embodies the 
exact asymptotic behaviors at both small and large wave number $q$. 
This allows us to also provide a closed-form expression for the exchange 
and correlation kernel $K_{\rm \scriptscriptstyle xc}(r)$, which 
represents a key input for density functional studies of inhomogeneous
systems.
\end{abstract}
\vspace{0.5 cm}

The static charge-charge response function 
$\chi_{\rm \scriptscriptstyle C}(q)$ of a paramagnetic electron gas (EG)
can be written in terms of the Lindhard function $\chi_0(q)$ by means of
the spin-symmetric many-body local field $G_+(q)$ through the relationship
\begin{equation}
\chi_{\rm \scriptscriptstyle C}(q)
=\frac{\chi_0(q)}{1-v_{q}[1-G_+(q)] \chi_0(q)}.
\end{equation}
Thus $G_+(q)$ is a fundamental quantity for the 
determination of many properties of a general electron system.
By definition $G_+(q)$ is meant to represent the effects of the 
exchange and correlation hole surrounding each electron in the fluid and
is therefore a key input in the density functional theory (DFT) of the 
inhomogeneous electron gas\cite{gross} and in studies of 
quasiparticle properties (such as the effective mass and 
the effective Land\`e g-factor) in the electronic Fermi
liquid\cite{giuliani}.  

For what concerns DFT calculations, a common approximation to the unknown
exchange-correlation energy functional $E_{\rm \scriptscriptstyle
xc}[n]$ appeals to its 
second functional derivative, 
\begin{equation}
K_{\rm \scriptscriptstyle xc}({\bar n}, |{\bf r}-{\bf r}'|)\equiv
\left.\frac{\delta^{2} E_{\rm
\scriptscriptstyle xc}[n]}{\delta n({\bf r}) \delta n({\bf
r}')}\right|_{{\bar n}}\, ,
\end{equation}
where ${\bar n}$ is the average local density of the EG. The local
field factor and 
the exchange-correlation kernel are simply related in Fourier
transform by
\begin{equation}\label{1}
{\widetilde K}_{\rm \scriptscriptstyle xc}(q)\equiv \int \!d^{d}r\,e^{-i\,{\bf
q}\cdot {\bf r}}\,K_{\rm \scriptscriptstyle xc}(r) =-v_{q}\,G_{+}(q)\, ,
\end{equation}
where $d$ is the dimensionality of the system and 
$v_{q}$ is the Fourier transform of the Coulomb potential
$e^{2}/r$. In what follows we shall only consider the case of
two spatial dimensions, with $d=2$ and $v_{q}=2 \pi e^{2}/q$. 
The corresponding three-dimensional case was discussed in
Ref.~[$\!\!$\onlinecite{Corradini}].  

A number of exact asymptotic properties of the static local field
factor in two dimensions are readily proven. In particular,

\begin{equation}\label{smallq}
\lim_{q \rightarrow 0}~G_{+}(q)= A_+ \frac{q}{k_{F}}
\end{equation}
with 
\begin{equation}
A_+=\frac{1}{r_{s}\sqrt{2}}\,\left(1-\frac{\kappa_{0}}{\kappa}\right)~,
\end{equation}
where $k_{F}=\sqrt{2 \pi n}=\sqrt{2}/r_{s} a_{B}$ is the Fermi wave
number, $r_{s}=\sqrt{\pi n a^{2}_B}$ is the usual EG density parameter
with $a_{B}$ the Bohr radius, $\kappa_{0}=\pi
r^{4}_{s}/2$ is the compressibility of the
ideal gas in units of $a^{2}_{B}/{\rm Ryd}$, 
while $\kappa$ is the compressibility of the interacting system.
By making use of the thermodynamic definition of $\kappa$ we can write
\begin{equation}
\frac{\kappa_{0}}{\kappa}=1-\frac{\sqrt{2}}{\pi}\,r_{s}+\frac{r^{4}_{s}}{8}
\left[\frac{d^{2} \epsilon_{c}(r_{s})}{d r^{2}_{s}}-\frac{1}{r_{s}}
\frac{d \epsilon_{c}(r_{s})}{d r_{s}}\right] \, ,
\end{equation}
where $\epsilon_{c}(r_{s})$ is the correlation energy per particle. 
Once this function is known, it is possible to calculate $A_+$. 
For the present purpose
$\epsilon_{c}(r_{s})$ can be taken from the Monte Carlo data of 
Ref.~[$\!\!$\onlinecite{rap}].

The asymptotic behavior of $G_{+}(q)$ at large $q$ is also known
exactly\cite{SanGiu,holas}:  
\begin{equation}\label{largeq}
\lim_{q\rightarrow \infty}
G_{+}(q)=C_+\, \frac{q}{k_{F}}+B_+~,
\end{equation}
where $C_+$ is proportional to the difference in kinetic energy between the
interacting and the ideal gas,
\begin{equation}
C_+=\frac{t-t_{0}}{2 \pi n e^{2}}\,k_{F}= 
-\frac{r_{s}}{2\sqrt{2}}\,\frac{d}{d
r_{s}}\Big(r_{s}\epsilon_{c}(r_{s})\Big)~.
\end{equation}
Moreover $B_+=1-g(0)$, $g(0)$ being the value of the pair-correlation
function at the origin. For $g(0)$ we use the simple expression
\begin{equation}
g(0)=\frac{1/2}{1+1.372\,r_{s}+0.0830\,r^{2}_{s}},
\end{equation}
which has been derived\cite{marco} by an interpolation between the result of a
low-$r_{s}$ expansion, including the second order direct and exchange
contributions to the energy in the paramagnetic state, and the result
of a partial-wave phase-shift analysis near Wigner
crystallization. This interpolation formula is in excellent agreement
with many-body calculations based on the ladder
approximation\cite{free,nagano}. 

In this work we fit the values of $G_{+}(q)$ originally obtained by
Diffusion-Monte-Carlo (DMC) in Ref.~[$\!\!$\onlinecite{MCS}]
in such a way as to obtain analytical expressions for both 
${\widetilde K}_{\rm \scriptscriptstyle xc}(q)$ and 
$K_{\rm \scriptscriptstyle xc}(r)$. 
Our formula for $G_{+}(q)$ reads
\begin{equation}\label{fit}
G_{+}({\bar q})=A_+\,{\bar q}\,\,
\left[\frac{e^{r_s/10}}{\sqrt{1+\left(A_+\,e^{r_s/10}\,{\bar
q}/B_+\right)^{2}}}+\Big(1-e^{r_s/10}\Big)\,
e^{-{\bar q}^{2}/4}\right] +
C_+\,{\bar q}\left(1-e^{-{\bar q}^{2}}\right)+{\rm P}_+({\bar q})\,
e^{-\alpha_+\,{\bar q}^{2}}\, ,
\end{equation}
where ${\bar q}=q/k_{F}$  and ${\rm P}_+({\bar q})$ is the polynomial
${\rm P}_+({\bar q})=g_2\,{\bar q}^{2}+g_4\,{\bar q}^{4}+g_6\,{\bar
q}^{6}+ g_8 \, {\bar q}^{8}$.

Some comments are needed in order to appreciate the correct physics
which has been incorporated in Eq.~(\ref{fit}). (i) Our functional 
form embodies the exact asymptotic behaviors
already introduced in Eq.~(\ref{smallq}) and (\ref{largeq}).
(ii) The exponential factor $e^{r_s/10}$ ensures that $G_+(q)$ rapidly
reaches the asymptotic behavior given by Eq.~(\ref{largeq}), a fact
that is borne out by the DMC data at $r_s=10$.
(iii) In the high-density limit ($r_s\rightarrow 0$) the term in
square brackets tends to a two-dimensional Hubbard-like
term\cite{jons}, while the second and third terms tend to zero.
(iv) The introduction of the high-degree polynomial ${\rm P}_+(q)$ 
serves to reproduce the rich structure at intermediate wave number
which is exhibited by $G_+(q)$ as compared to the three-dimensional
case\cite{MCS3}. 

\begin{figure}[h!]
\centerline{\mbox{\psfig{figure=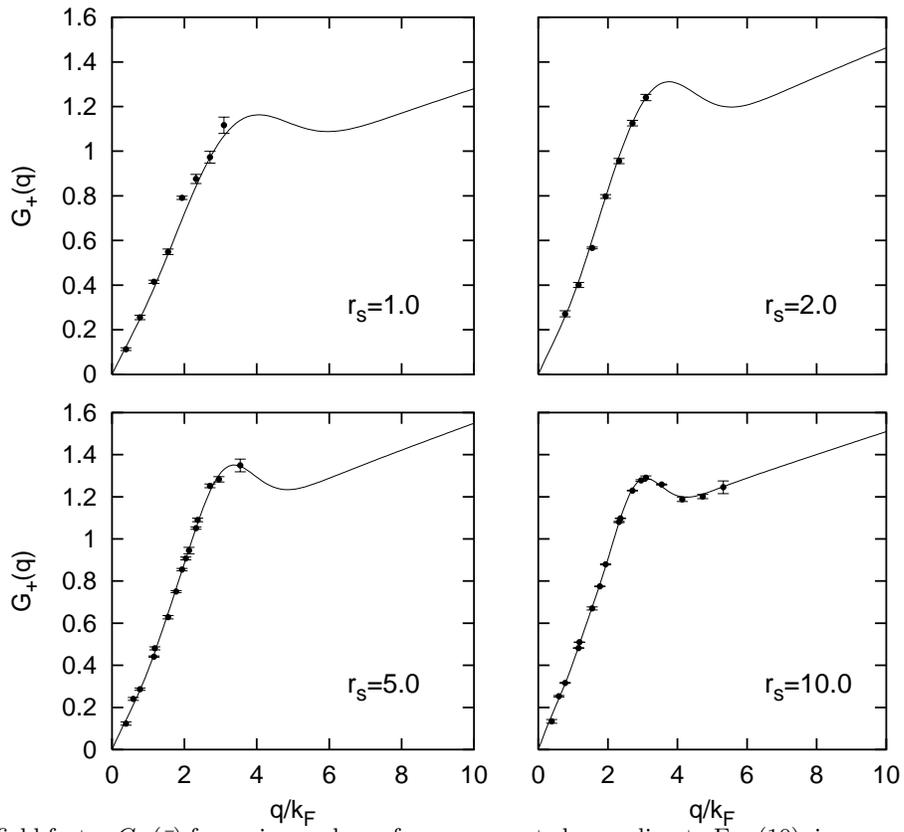, angle =0, width = 12 cm}}}
\caption{The local field factor $G_{+}({\bar q})$ for various values
of $r_{s}$ as computed according to Eq.~(\ref{fit}), in comparison
with the DMC data of Ref. 10.}
\label{fig1}
\end{figure}

\begin{figure}[h!]
\centerline{\mbox{\psfig{figure=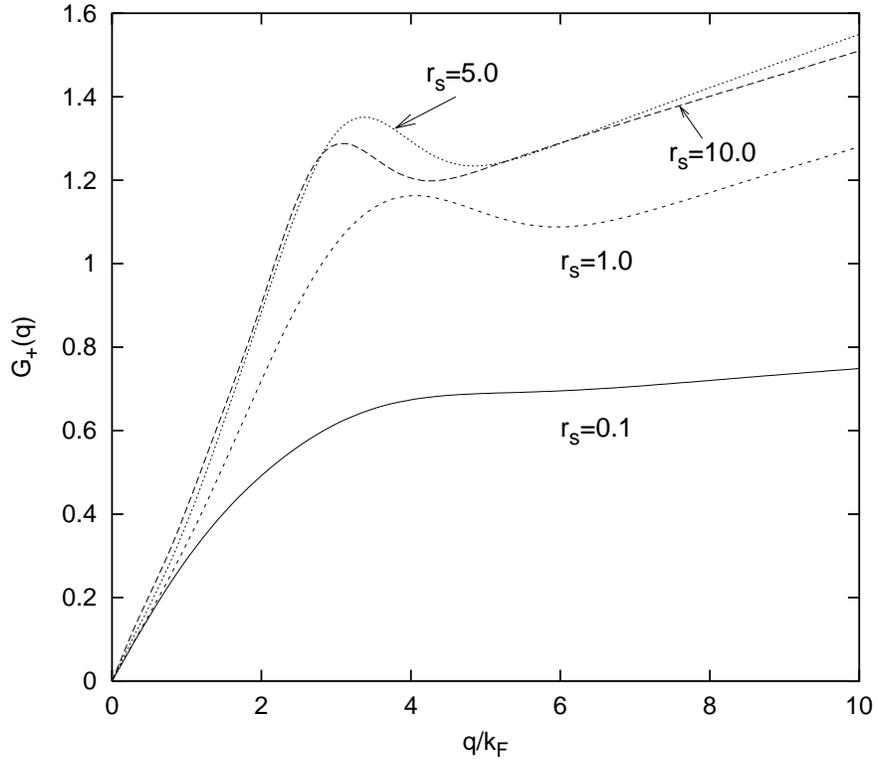, angle =0, width = 12 cm}}}
\caption{The local field factor $G_{+}(q)$ as from Eq. (\ref{fit})
for various values of $r_{s}$.}
\label{fig2}
\end{figure}

The only free parameters are contained in the last term in Eq. (\ref{fit}) 
and are fitted so as to minimize the differences from the DMC
numerical results.
For practical reasons it proves useful to have a continuous 
parametrization  of the coefficients of the polynomial ${\rm P}_+(q)$,
which is at least valid in the range $0\leq r_{s} \leq 10$.
We therefore propose the following:
\begin{eqnarray}\label{fitta}
&&\alpha_+(r_{s})=\frac{0.1598+0.8931\,(r_{s}/10)^{0.9218}}
{1+0.8793\,(r_{s}/10)^{0.9218}}~,\nonumber \\
&& g_2 (r_{s})= 0.5824\,(r_{s}/10)^2-0.4272\,(r_{s}/10)~,\nonumber\\
&& g_4
(r_{s})=0.2960\,(r_{s}/10)-1.003\,(r_{s}/10)^{5/2}+0.9466\,(r_{s}/10)^3~, 
\nonumber \\
&& g_6 (r_{s})=-0.0585\,(r_{s}/10)^{2}~,\nonumber \\
&& g_8 (r_{s})=0.0131\,(r_{s}/10)^2\, . 
\end{eqnarray}
In Figure~\ref{fig1} we compare the fit given by 
Eq.~(\ref{fit}) and (\ref{fitta}) with the DMC data for $r_{s}=1,2,5$ and $10$. 
In Figure~\ref{fig2} we show the local field factor $G_{+}(q)$ as from
Eq.~(\ref{fit}) for various values of $r_{s}$: the evolution from the
low-$r_{s}$ regime to the high-$r_{s}$ one is clear. The fact that 
the highest peak in $G_{+}(q)$ occurs at $r_{s}=5$ is due to the
behavior of $C_+(r_{s})$: this is a function that increases up to
$r_{s}\simeq 3.5$, reaches a maximum and then decreases. 
Thus, the value of $C_+$ at $r_{s}=5$ is larger than that at $r_{s}=10$. 

We turn next to the evaluation of $K_{\rm \scriptscriptstyle xc}(r)$. 
>From Eq.~(\ref{1}) and (\ref{fit}) the expression of the exchange-correlation
kernel in real space
(in Ryd) is readily obtained as
\begin{equation}\label{kxc}
K_{\rm \scriptscriptstyle xc}(r) = M_1 \frac{\delta^{(2)}({\bf r})}{k_F^2} 
+ M_2\,\frac{\exp{\Big(-B_+\,k_Fr/(A_+e^{r_{s}/10})\Big)}}{k_Fr}+
M_3\,e^{-(k_Fr)^2}+M_4\,e^{-(k_Fr)^2/4}+
\sum_{n=1}^{4} M_{5,2n} F_{2n} (\alpha_+, k_Fr) ~,
\end{equation}
where $M_{1}=-4\,\pi\, \sqrt{2}\, C_+/r_{s}$, 
$M_{2}=-2\sqrt{2}\,B_+/r_{s}$, 
$M_3=-4\,\sqrt{2}\,A_+\,(1-e^{r_{s}/10})/r_{s}$,
$M_4=\sqrt{2}\,C_+/r_{s}$ and $M_{5,n} = -2^\frac{3}{2}\,g_{n}\,/r_s$.
The function $F_n(\alpha, x)$ is given by
\begin{equation}
F_n (\alpha, x)=\int_{0}^{\infty}\,dy\,y^{n}\,{\rm
J}_{0}(xy)\,e^{-\alpha\,y^2}=\frac{1}{2}
\,\alpha^{-(1+n)/2}\,\Gamma(\frac{1+n}{2})\,\,_{1}{\rm
F}_{1}\left(\frac{1+n}{2};1; -\frac{x^2}{4\alpha}\right) 
\end{equation}
where $\Gamma(z)$ is Euler's Gamma function and 
$_{1}{\rm F}_{1}(a;b;z)$ is Kummer's function. 
In practice the function $F_{n}(\alpha, x)$ 
can be obtained via the recursive relation
\begin{equation}
F_{n+2}(\alpha, x)=-\frac{d \,F_{n}(\alpha, x)}{d\,
\alpha}~,\,\,\,\,F_{2}(\alpha,x)=\frac{\sqrt{\pi}}{16\,
\alpha^{5/2}}\,\left[(4\alpha-x^2)\,{\rm
I}_{0}\left(\frac{x^{2}}{8\alpha}\right)+x^2\, {\rm
I}_{1}\left(\frac{x^{2}}{8\alpha}\right)\right]\,e^{-x^{2}/8\alpha}
\end{equation}
where ${\rm I}_{n}(z)$ is the modified Bessel function of order
$n$. It is also useful to recall that $d {\rm I}_{0}(z)/dz={\rm
I}_{1}(z)$ and that $d {\rm I}_{1}(z)/dz={\rm I}_{0}(z)-{\rm
I}_{1}(z)/z$.

\begin{figure}[h!]
\centerline{\mbox{\psfig{figure=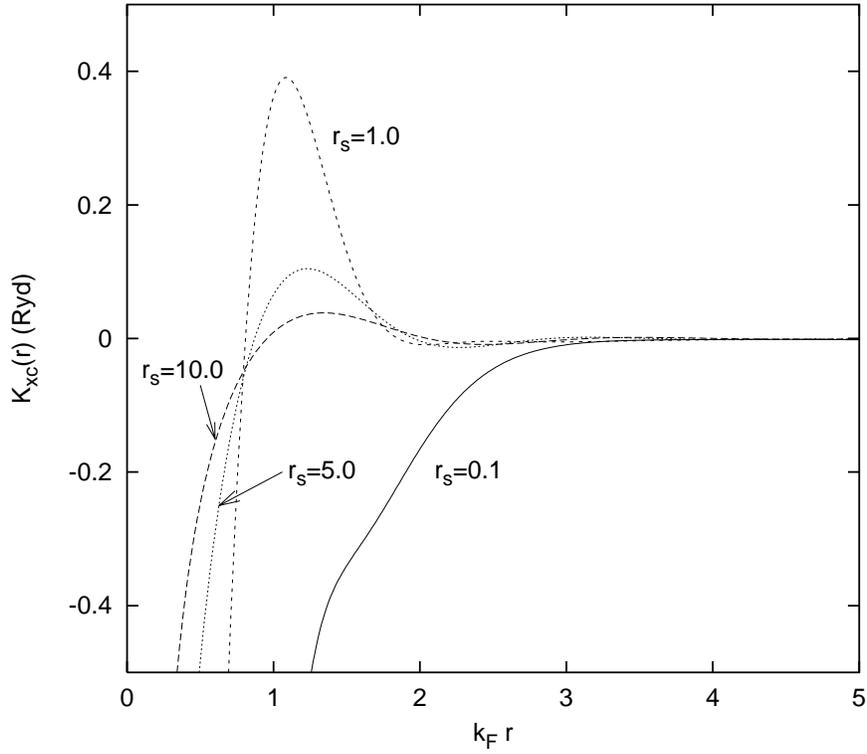, angle =0, width = 12 cm}}}
\caption{The exchange-correlation kernel $K_{xc}(r)$ 
as from Eq.~(\ref{kxc})
for various values of $r_{s}$.}
\label{fig2bis}
\end{figure}

In Figure~\ref{fig2bis} we show the exchange-correlation kernel 
$K_{\rm \scriptscriptstyle xc}(r)$ for various values of $r_{s}$ as from
Eq.~(\ref{kxc}) (without the first term, which contains a
two-dimensional $\delta$ function). It
is pleasing to note that no long-range oscillations are present in 
$K_{\rm \scriptscriptstyle xc}(r)$. Notice that the $M_2$ term in 
Eq.~(\ref{kxc}) diverges for $k_Fr \rightarrow 0$.

In conclusion, we have presented an analytic parametrization of the
local field factor entering the dielectric response of the two-dimensional
electron gas in the paramagnetic state, incorporating the known
asymptotic behaviors and giving an accurate description of the
available quantum Monte Carlo data. We have obtained from it an
analytic expression of the exchange-correlation kernel for
density-functional calculations on inhomogeneous two-dimensional
electronic systems.
\acknowledgements
This work was partially supported by MURST through the PRIN 1999 program.
We are grateful to Dr. S. Moroni for fruitful 
discussions and for providing us 
with the results of the Diffusion Monte Carlo study. 


\begin{references}
\bibitem{gross} R. M. Dreizler and E. K. U.  Gross, {\it
Density Functional Theory, An Approach to the Quantum Many-Body Problem} 
(Springer, Berlin 1990).
\bibitem{giuliani}
S. Yarlagadda and G. F. Giuliani, Phys. Rev. B {\bf 40}, 5432 (1989);
S. Yarlagadda and G. F. Giuliani, Phys. Rev. B {\bf 49}, 14188 (1994).
\bibitem{Corradini}
M. Corradini, R. Del Sole, G. Onida and M. Palummo,
Phys. Rev. B {\bf 57}, 14569 (1998).
\bibitem{rap}
F. Rapisarda and G. Senatore, Aust. J. Phys. {\bf 49}, 161 (1996).
\bibitem{SanGiu}
G. E. Santoro and G. F. Giuliani, Phys. Rev. B {\bf 37}, 4813 (1988).
\bibitem{holas}  A. Holas, in {\it Strongly Coupled Plasma Physics},
eds. F. J. Rogers and  H. E. DeWitt, (Plenum, New York 1986), p. 463.
\bibitem{marco}
M. Polini, G. Sica, B. Davoudi and M. P. Tosi, J. Phys.:
Condens. Matter. {\bf 13}, 3591 (2001).
\bibitem{free}
D. L. Freeman, J. Phys. C {\bf 16}, 711 (1983).
\bibitem{nagano}
S. Nagano, K. S. Singwi and S. Ohnishi, Phys. Rev. B {\bf 29}, 1209
(1984).
\bibitem{MCS}
S. Moroni, D. M. Ceperley and G. Senatore,
Phys. Rev. Lett. {\bf 69}, 1837 (1992); G. Senatore, S. Moroni and
D. M. Ceperley, in {\it Quantum Monte Carlo Methods in Physics and
Chemistry}, eds. M. P. Nightingale and C. J. Umrigar (Kluwer,
Dordrecht 1999); and private communication.
\bibitem{jons}
M. Jonson, J. Phys. C {\bf 9}, 3055 (1976); see also
Ref.~[$\!\!$\onlinecite{SanGiu}].
\bibitem{MCS3} S. Moroni, D. M. Ceperley and G. Senatore,
Phys. Rev. Lett. {\bf 75}, 689 (1995).
\end{references}
\end{document}